# MEASUREMENT OF THE TIME STRUCTURE OF FLASH BEAMS USING PROMPT GAMMA RAYS AND SECONDARY NEUTRONS AS SURROGATES

Serdar Charyyev[1,5], Ruirui Liu[2], Xiaofeng Yang[2], Jun Zhou[2], Anees Dhabaan[2], William S. Dynan[2,3], Cristina Oancea[4] and Liyong Lin[2]

[1] Department of Radiation Oncology, Stanford University, Palo Alto, CA 94305, United States

[2] Department of Radiation Oncology and Winship Cancer Institute, Emory University, Atlanta, GA 30322, United States

[3] Department of Biochemistry, Emory University School of Medicine, Emory University, Atlanta, GA 30322, United States

[4] ADVACAM, Prague, Czech Republic

[5] Author to whom any correspondence should be addressed.

Email: charyyev@stanford.edu

## Abstract

We aim to investigate the feasibility of online monitoring of irradiation time (IRT) and scan time for FLASH radiotherapy using a pixelated semiconductor detector.

Measurements of the time structure of FLASH irradiations were performed using fast, pixelated spectral detectors, AdvaPIX-TPX3 and Minipix-TPX3. The latter has a fraction of its sensor coated with a neutron sensitive material. With little or no dead time and an ability to resolve events that are closely spaced in time (tens of ns), both detectors can accurately determine IRTs as long as pile-ups are avoided. To avoid pile-ups, we placed the detectors beyond the Bragg peak or at a large scattering angle. We acquired prompt gamma rays and secondary neutrons and calculated IRTs based on timestamps of the first (beam-on) and the last (beam-off) charged species. We also measured scan times in x, y, and diagonal directions. We performed these measurements for a single spot, a small animal field, a patient field, and a ridge filter optimized field to demonstrate *in vivo* online monitoring of IRT. All measurements were compared to vendor log files.

Differences between measurements and log files for a single spot, a small animal field, and a patient field were within 1%, 0.3% and 1%, respectively. *In vivo* monitoring of IRTs was accurate within 0.1% for AdvaPIX-TPX3 and within 6.1% for Minipix-TPX3. The scan times in x, y, and diagonal directions were 4.0, 3.4, and 4.0 ms, respectively.

Overall, the AdvaPIX-TPX3 can measure FLASH IRTs within 1% accuracy, indicating that prompt gamma rays are a good surrogate for primary protons. The Minipix-TPX3 showed a higher discrepancy, suggesting a need for further investigation. The scan times (3.4 ± 0.05 ms) in the 60-mm distance of y-direction were less than (4.0 ± 0.06 ms) in the 24-mm distance of x-direction, confirming the much faster scanning speed of the Y magnets than that of X.

1. Introduction

A series of studies (Favaudon *et al.*, 2014; Loo *et al.*, 2017; Vozenin *et al.*, 2019a; Montay-Gruel *et al.*, 2017) have demonstrated that irradiation delivered with ultra-high dose rate (>40 Gy/s), also known as FLASH, has a sparing effect on normal tissue due to 'FLASH effect'. In their groundbreaking work, Favaudon et al. (Favaudon *et al.*, 2014) have demonstrated the effect in nude mice in which they used low-energy electrons to irradiate their target. The same FLASH irradiator was used to treat the first patient with FLASH radiotherapy (RT), a cutaneous lymphoma patient (Bourhis *et al.*, 2019). Utility of electrons is limited to the superficial targets as they are not penetrative enough for deeper tumors.

There are efforts that have shown FLASH effect is possible with x-rays, both in kV (Montay-Gruel *et al.*, 2019; Montay-Gruel *et al.*, 2018; Smyth *et al.*, 2018) and MV (Kutsaev *et al.*, 2021) range. Both kV and MV x-rays are suitable for irradiation of small volumes and for in vitro studies; however, isocentric treatments which achieve FLASH dose rates are difficult to realize.

An attractive and readily available option to deliver a FLASH beam is with cyclotron-based protons, specifically pencil beam scanning (PBS), since PBS can be better controlled in terms of position and intensity (Paganetti, 2012). Not only that, but also protons are more penetrative, offer finite range, and are radiobiologically more effective (Paganetti *et al.*, 2019). Only minor adjustments to clinical beam parameters are needed to make existing proton therapy facilities 'FLASH-ready'.

Cyclotron-based proton systems have pulse repetition and duration rates on the order of nanoseconds and can be considered as continuous wave sources (Ashraf *et al.*, 2020). Dose rate for FLASH studies is defined as $\frac{delivered\ dose}{IRT}$, i.e., average dose, requiring accurate measurement of both the dose and the IRT. Usually, IRT is obtained from log files (Diffenderfer *et al.*, 2021) and needs to be verified with measurements. One way to measure IRT is to use prompt gamma rays (PG) or secondary neutrons (SN) which are emitted within nanoseconds after proton interactions, which can be considered 'real-time' (Golnik *et al.*, 2014; Diffenderfer *et al.*, 2020; Haertter *et al.*, 2021). Consequently, they are good surrogates for primary proton IRT. Haertter et al. investigated the feasibility of real-time dose rate monitoring via PG timing using NaI(Tl) crystal coupled to a photomultiplier tube. They have concluded that due to scintillation afterglow, real-time dose monitoring is possible for irradiations less than 36 Gy/s. For real-time monitoring with higher dose rates, NaI is susceptible to PG count saturation (Haertter *et al.*, 2021). More recently, strip ionization chambers are becoming popular for real-time monitoring (Zhou *et al.*, 2022; Yang *et al.*). Yang *et al.* successfully demonstrated the feasibility of FLASH beam monitoring up to a 20-kHz sampling rate using a 2D strip ionization chamber.

As an alternative and potentially 100-1000 times faster, a semiconductor detector can be used for the better IRT monitoring purposes, especially in a radiofrequency modulated beams with ~10 ns pulses. Also, the high spatial resolution, 55 µm, of the detector proposed in our method makes it possible to perform linear energy transfer (LET) measurements. In this work, we aim to investigate the feasibility of online monitoring of IRT and scan time for FLASH radiotherapy using a pixelated semiconductor detector.

2. Materials and Methods

*Detector and setup conditions*

We tracked PG and SNs using fast, pixelated spectral detectors, AdvaPIX-TPX3 and Minipix-TPX3. The latter has a fraction of its sensor coated with a material to increase sensitivity to neutrons. A detailed description of each detector and their capabilities are explained elsewhere (Granja *et al.*, 2018; Granja *et al.*, 2021; Charyyev *et al.*, 2021). Briefly, each comes with an advanced semiconductor pixel ASIC readout chip, Timepix3, bonded to either 500 μm (for AdvaPIX-TPX3) or 600 μm (Minipix-TPX3) thick silicon. The time-of-arrival can be identified with a precision of 1.56 ns (Granja *et al.*, 2018).

To avoid pile-ups, we placed the AdvaPIX-TPX3 well beyond Bragg peak where gamma fluence is low, and we placed Minipix-TPX3 at a large scattering angle to detect SNs, Figure 1a. In Figure 1b, one can see the fluence depth distribution for protons (blue curve) and resulting PG (red curve). We obtained these fluence depth distributions using a well-established Monte Carlo code, Tool for Particle Simulation (TOPAS) (Perl *et al.*, 2012), version 3.1.p2. Modeling of PBS characteristics in TOPAS is done based on methods described elsewhere (Charyyev *et al.*, 2020). At detector location, PG fluence is two orders of magnitude less, explaining the rationale of placing the detector beyond the Bragg peak. Figure 1c shows the setup of anthropomorphic phantom to demonstrate the feasibility of in vivo online monitoring.

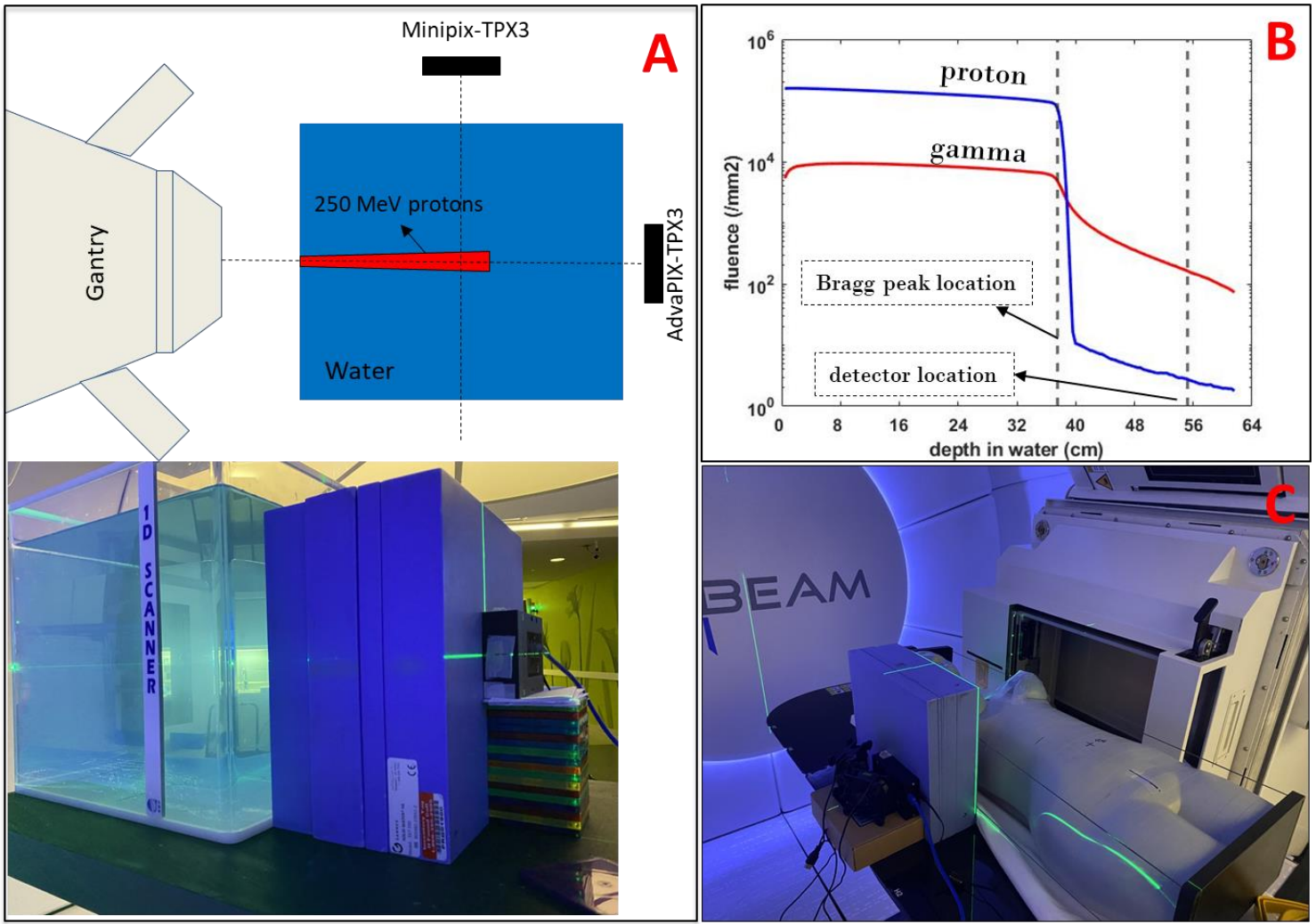

**Figure 1.** *(a) Bird's-eye view of the placement of the detectors. (b) Fluence depth distribution for protons (blue curve) and resulting PG (red curve). Y-axis is in log scale and demonstrates the rationale of why we placed detector beyond Bragg peak. (c) Setup of anthropomorphic phantom to demonstrate the feasibility of in vivo online monitoring.*

### *Principle of IRT determination*

Each incident particle generates a charge signal that is detected in the sensor and time-stamped. Based on these charges, IRT can be calculated based on timestamps of the first charged species (beam-on) and the last charged species (beam-off) arriving on the sensor. Figure 2 shows the principle of IRT determination on a sample measured FLASH field, with an average dose rate (Folkerts *et al.*, 2020) of ~50 Gy/s delivered over 270 ms. Panel (a) illustrates particle detection and track visualization for a photon, an electron, a proton and a heavy charged particle with each of them having a distinct morphology. Panel (b) lists first 30 of 150,000 total particles registered by the detector for this sample measurement. Panel (c) shows a plot of the time structure of this sample FLASH field. The beam is on during the steep portion of the curve. Panel (d) shows time structure on an expanded scale, corresponding to 1.5 µs of the 270 ms irradiation.

With little or no dead time and an ability to resolve events that are closely spaced in time (tens of nanoseconds), both detectors can accurately determine IRTs as long as pile-ups are avoided (Usman and Patil, 2018).

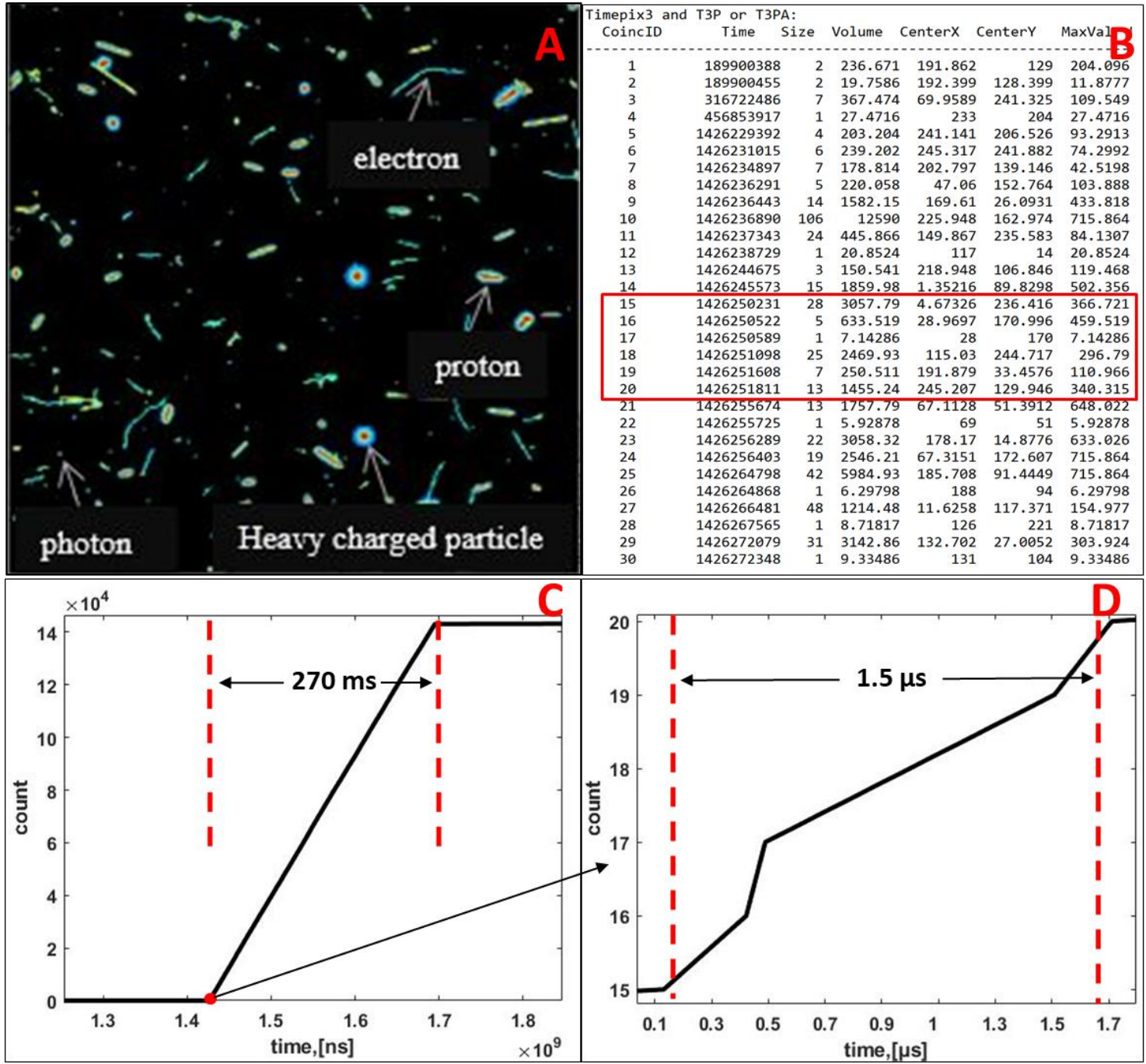

**Figure 2.** *(a) Detection and track visualization for different species of particles. (b) Particles as registered by the detector from this sample FLASH field measurement (only the first 30 registrations are shown). (c) Time structure of this FLASH field. (d) Magnified section from (c) to show the particles 15-20 from (b), shown in red rectangle, or with a red dot in (c).*

*Workflow and other measurement details*

Our PBS system, Varian ProBeam, is capable of delivering 250 MeV (the highest energy) at various beam currents in its special 'Racehorse' mode. We followed the steps as outlined in Figure 3 to demonstrate accurate IRT and scan time measurements and feasibility of online time monitoring. We began with a measurement of a single spot IRT at clinical beam current, ~5 nA. We gradually increased the beam current to test if and when the IRT measurement becomes inaccurate, thereby exposing the detector sensor to a gradual step up of the flux and testing its limits at FLASH dose rates. This step was followed by a measurement of IRT of a small animal field, a 30x30 mm$^2$ field at isocenter created using scanned spots with 5 mm spot spacing (i.e., 7 spots in both x- and y-direction, with a total of 49 spots) for irradiating mice as part of an ongoing small animal studies in our center. Measurement for the small animal field was repeated five times to check reproducibility and to quantify the uncertainty. We then measured scan times and speeds in x, y, and diagonal directions. To do this, we delivered two spots separated by 24 mm in the x-direction or 60 mm in the y-direction. For the diagonal scan, we delivered two spots separated by 24 mm in both x- and y- directions. We also demonstrated the capability of the detector to measure the beam-off by triggering a beam fault when trying to scan in x-direction with a very large insterspot distance of 60 mm. Next, we measured the IRT of a patient field with spots of irregular spacings to mimic realistic fields that could be delivered in clinical scenarios. Lastly, we delivered ridge filter optimized field and measured IRT for it, first with AdvaPIX-TPX3 followed by a measurement with Minipix-TPX3. We performed these measurements using anthropomorphic phantom to demonstrate the feasibility of *in vivo* online monitoring. The ridge filter optimized field has irregular spot spacings and irregular monitor units for each spot and was designed based on a real patient's tumor target and anatomy information. It is optimized to cover 3 cm diameter tumor volume using 250 MeV proton beam (Liu *et al.*, 2021; Liu *et al.*, 2022).

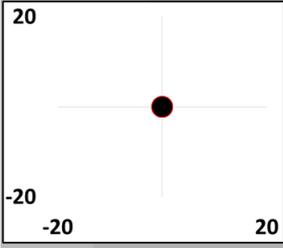

**Figure 3.** *Measurements steps followed to demonstrate IRT measurements and feasibility of online monitoring of time. Spot maps are provided for all irradiations described.*

3. **Results**

Figure 4a illustrates measured single spot IRTs as a function of beam current. IRT decreases as the beam current is increased for the first three increments and then stabilizes at 20 nA beam current. We measured IRT around constant 48 ms for the beam currents 20-150 nA. Our detector accurately measured IRT for beam current of 150 nA, highest our machine can achieve, without further modifications. Figure 4b illustrates the IRT reproducibility for a small animal field, a 30x30 mm$^2$ at isocenter created using scanned spots with 5 mm spot spacing (spot map is provided in Figure 3). The minimum, maximum and average values of IRT are 287, 289 and 288.16 ms, respectively, and the standard deviation of the five measurements is 0.74 ms. This small animal field delivers 17.5 Gy in 288 ms, thereby demonstrating accurate IRT measurements at 60.7 Gy/s average dose rate.

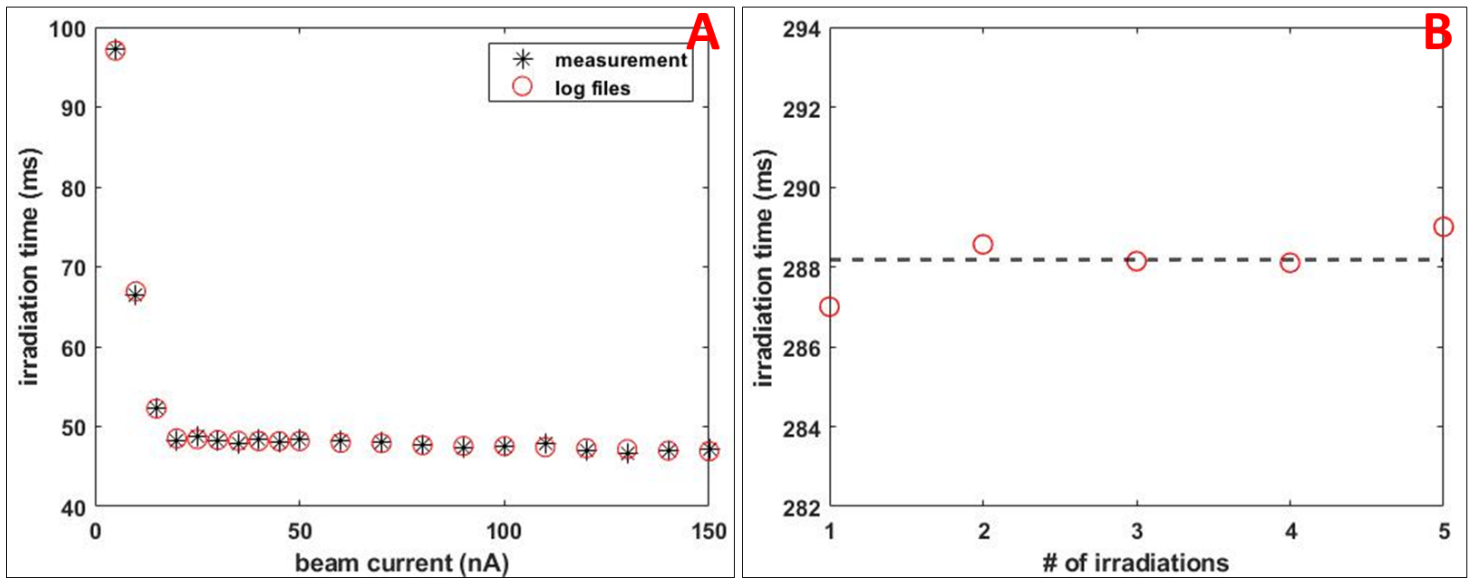

**Figure 4. (A)** *Measured single spot IRTs as a function of beam current.* **(B)** *The IRT reproducibility for a small animal field, a 30x30 mm$^2$ at isocenter created using scanned spots with 5 mm spot spacing.*

On average, patient field IRT was measured within 1% of log files. *In vivo* demonstration of IRT measurement was within 0.1% of log files using AdvaPIX-TPX3 (using PGs as a surrogate) and within 6.1% of log files using Timepix-TPX3 (using SNs as a surrogate). In Table 1, we summarize our measurements and compare with log files obtained from the vendor. Values provided in Table 1 are obtained from repeated measurements: 17 times (standard deviation ±0.6 ms) for the single spot and 5 times (standard deviation ±0.7 ms) for the small animal field. Measurements are repeated twice for the rest of the scenarios.

**Table 1.** *Comparison of our measurements with log files obtained from the vendor for different FLASH irradiations scenarios investigated in this work. Reported values are average values of repeated measurements.*

| Description | Measurement Time (ms) | Log File Time (ms) | % [min, max] difference |
|---|---|---|---|
| Single Spot | 47.788 | 47.786 | [-0.970, 0.820] |
| Small Animal Field (3x3cm$^2$) | 288.16 | 288.28 | [-0.34, 0.19] |
| Patient Field | 1920.7 | 1902.5 | [0.4, 0.9] |
| In-vivo, ADV-TPX3, PGs as surrogate | 764.74 | 764.75 | [<-0.01, 0.04] |
| In-vivo, Mini-TPX3, SNs as surrogate | 808.5 | 764.7 | [5.6, 6.1] |

Figure 5 shows scan times in x, y, and diagonal directions, as reflected by the plateau region of the time structure for each irradiation. Each plot encompasses the irradiation of two spots, separated by 24 mm in the x-direction (Figure 5A), 60 mm in the y-direction (Figure 5B), or 24 mm in both the x- and y-directions (Figure 5C, diagonal scan). The scan time in the x-direction, 4 ± 0.06 ms, resulted in scan speed of 6 mm/ms and the scan time in the y-direction, 3.37 ± 0.05 ms, resulted in scan speed of 17.8 mm/ms, confirming the faster scanning speed of the Y magnets. Diagonal scan speed was limited by the slower X magnets. Figure 5D shows the time structure of the beam fault that was triggered when two spots were separated by 60 mm and scanned in x-direction.

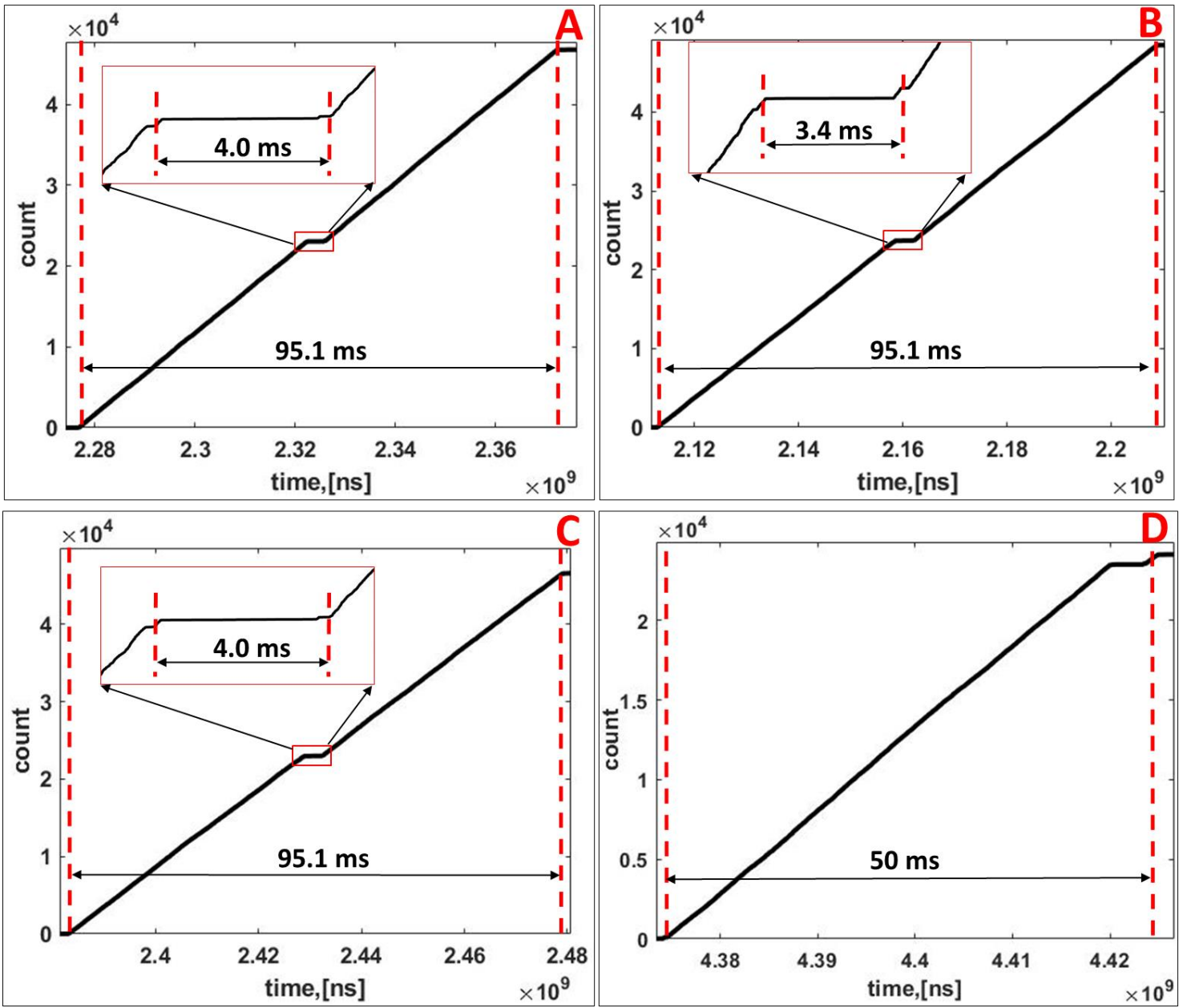

**Figure 5.** *Time structure of irradiation of two spots separated by (A) 24 mm and scanned in x-direction, (B) 60 mm and scanned in y-direction, (C) 24 mm in both the x- and y-directions (diagonal scan). (D) shows the time structure of the beam fault that was triggered when two spots were separated by 60 mm and scanned in x-direction.*

4. Discussion

Among the results presented in Section 3, several observations are noted and discussed in this section. First, one would expect IRT to get shorter and be linear, in theory, as beam current is increased. Looking at Figure 4a, we can see that it is getting shorter (though not perfectly linear) up to 15 nA beam current and then stays constant up to 150 nA beam current. We have observed that beyond 15 nA, monitoring ionization chamber saturates (probably due to ion recombination issues) and does not shut off beam on time. We have shown that independent raw charge measurements increase as a function of increasing beam current beyond 15 nA (Charyyev et al., 2022).

Dose rate for most FLASH studies is reported in terms of average dose rate, i.e., $\frac{delivered\ dose}{IRT}$, which was the definition we adopted in this study. Even though there is a consensus that average dose rate is the most relevant for FLASH effect (Vozenin *et al.*, 2019b; Wilson *et al.*, 2019; Petersson *et al.*, 2020), for proton PBS, dose rate can be defined, at each point in the field, as the sum of contributions from multiple spots (Folkerts *et al.*, 2020; Zou *et al.*, 2021). In that sense, considering the subset of voxels in a field, the dose rate can be higher that the dose rate reported in the Section 3 of this work. It is relevant to study dose rate in the context of sub-voxels, because the processes that are believed to be responsible for the FLASH effect happen at the cellular level.

In addition to being much faster way of online monitoring of IRT, the proposed method offers advantage of measuring LET over other existing methods (Charyyev *et al.*, 2021). Delivering the same LET radiation at FLASH dose rates will reduce oxygen enhancement ratio (OER) and increase relative biological effectiveness (RBE) (Jones, 2022). With this dependence of RBE/LET on LET, integrated biological optimizations frameworks are proposed (Liu *et al.*, 2022) where dose, dose rate and LET are optimized simultaneously to deliver a more effective FLASH irradiation. It is conceivable that there will be a need for a methodology to measure dose, dose rate and LET simultaneously. The extreme timing resolution (ns scale) would potentially be useful to characterize the timing structures of pulsed proton cyclotron and electron linear accelerators in the future besides the fine spatial resolution for LET measurements.

We can see that repeated measurements reveal delivery is reproducible within 1%. Because of instabilities of monitoring ionization chamber to control the beam, secondary and possibly tertiary confirmation measurements of dose rate have to be performed. Also, this uncertainty needs to be taken into account when interpreting outcomes of studies with current clinical system.

## 5. Conclusions

We have shown that IRT in order of few ms to order of few hundreds ms can be measured accurately with AdvaPIX-TPX3. The IRT measurements with Minipix-TPX3 showed a somewhat higher discrepancy, suggesting a need for further investigation when SNs are used as surrogates. Not only that, but also we were able to show scan time and differences between x-scan and y-scan speed in order of ms. Because our detector has a timing resolution in nanosecond scale, finer measurements are possible as long as pile-ups are avoided.

**Reference:**

Ashraf M R, Rahman M, Zhang R, Williams B B, Gladstone D J, Pogue B W and Bruza P 2020 Dosimetry for FLASH Radiotherapy: A Review of Tools and the Role of Radioluminescence and Cherenkov Emission *Frontiers in Physics* **8**

Bourhis J, Sozzi W J, Jorge P G, Gaide O, Bailat C, Duclos F, Patin D, Ozsahin M, Bochud F, Germond J F, Moeckli R and Vozenin M C 2019 Treatment of a first patient with FLASH-radiotherapy *Radiotherapy and oncology : journal of the European Society for Therapeutic Radiology and Oncology* **139** 18-22

Charyyev S, Artz M, Szalkowski G, Chang C W, Stanforth A, Lin L, Zhang R and Wang C C 2020 Optimization of hexagonal-pattern minibeams for spatially fractionated radiotherapy using proton beam scanning *Medical physics* **47** 3485-95

Charyyev S, Chang C-W, Zhu M, Lin L, Langen K and Dhabaan A 2022 Characterization of 250 MeV protons from Varian ProBeam pencil beam scanning system for FLASH radiation therapy.  p arXiv:2208.05037

Charyyev S, Chang C W, Harms J, Oancea C, Yoon S T, Yang X, Zhang T, Zhou J and Lin L 2021 A novel proton counting detector and method for the validation of tissue and implant material maps for Monte Carlo dose calculation *Phys Med Biol* **66** 045003

Diffenderfer E S, Sørensen B S, Mazal A and Carlson D J 2021 The current status of preclinical proton FLASH radiation and future directions *Medical physics*

Diffenderfer E S, Verginadis, II, Kim M M, Shoniyozov K, Velalopoulou A, Goia D, Putt M, Hagan S, Avery S, Teo K, Zou W, Lin A, Swisher-McClure S, Koch C, Kennedy A R, Minn A, Maity A, Busch T M, Dong L, Koumenis C, Metz J and Cengel K A 2020 Design, Implementation, and in Vivo Validation of a Novel Proton FLASH Radiation Therapy System *International journal of radiation oncology, biology, physics* **106** 440-8

Favaudon V, Caplier L, Monceau V, Pouzoulet F, Sayarath M, Fouillade C, Poupon M-F, Brito I, Hupé P, Bourhis J, Hall J, Fontaine J-J and Vozenin M-C 2014 Ultrahigh dose-rate FLASH irradiation increases the differential response between normal and tumor tissue in mice *Science translational medicine* **6** 245ra93-ra93

Folkerts M M, Abel E, Busold S, Perez J R, Krishnamurthi V and Ling C C 2020 A framework for defining FLASH dose rate for pencil beam scanning *Medical physics* **47** 6396-404

Golnik C, Hueso-González F, Müller A, Dendooven P, Enghardt W, Fiedler F, Kormoll T, Roemer K, Petzoldt J, Wagner A and Pausch G 2014 Range assessment in particle therapy based on prompt γ-ray timing measurements *Phys Med Biol* **59** 5399-422

Granja C, Kudela K, Jakubek J, Krist P, Chvatil D, Stursa J and Polansky S 2018 Directional detection of charged particles and cosmic rays with the miniaturized radiation camera MiniPIX Timepix *Nuclear Instruments and Methods in Physics Research Section A: Accelerators, Spectrometers, Detectors and Associated Equipment* **911** 142-52

Granja C, Oancea C, Jakubek J, Marek L, Benton E, Kodaira S, Miller J, Rucinski A, Gajewski J, Stasica P, Zach V, Stursa J, Chvatil D and Krist P 2021 Wide-range tracking and LET-spectra of energetic light and heavy charged particles *Nuclear Instruments and Methods in Physics Research Section A: Accelerators, Spectrometers, Detectors and Associated Equipment* **988** 164901

Haertter A, Kim M M, Zou J, Shoniyozov K, Avery S, Teo B, Metz J, Cengel K, Dong L and Diffenderfer E S 2021 Prompt Gamma Timing as a Real-Time Relative Dose Rate Monitor of FLASH Proton Delivery *Medical physics* **48**

Jones B 2022 The influence of hypoxia on LET and RBE relationships with implications for ultra-high dose rates and FLASH modelling *Physics in Medicine & Biology* **67** 125011

Kutsaev S V, Agustsson R, Arodzero A, Berry R, Boucher S, Diego A, Gavryushkin D, Hartzell J J, Lanza R C, Smirnov A Y, Verma A and Ziskin V 2021 Linear accelerator for security, industrial and medical applications with rapid beam parameter variation *Radiation Physics and Chemistry* **183** 109398

Liu R, Charyyev S, Wahl N, Liu W, Kang M, Zhou J, Yang X, Baltazar F, Palkowitsch M, Higgins K, Dynan W, Bradley J and Lin L 2022 An Integrated Biological Optimization framework for proton SBRT FLASH treatment planning allows dose, dose rate, and LET optimization using patient-specific ridge filters. p arXiv:2207.08016

Liu R, Charyyev S, Zhou J, Yang X, Liu T, McDonald M, Higgins K, Bradley J and Lin L 2021 Feasibility of 3D Printed Ridge Filter to Enable SBRT FLASH Therapy Using Scanning Proton Beam *Medical physics* **48**

Loo B W, Schuler E, Lartey F M, Rafat M, King G J, Trovati S, Koong A C and Maxim P G 2017 (P003) Delivery of Ultra-Rapid Flash Radiation Therapy and Demonstration of Normal Tissue Sparing After Abdominal Irradiation of Mice *International Journal of Radiation Oncology • Biology • Physics* **98** E16

Montay-Gruel P, Acharya M M, Petersson K, Alikhani L, Yakkala C, Allen B D, Ollivier J, Petit B, Jorge P G, Syage A R, Nguyen T A, Baddour A A D, Lu C, Singh P, Moeckli R, Bochud F, Germond J F, Froidevaux P, Bailat C, Bourhis J, Vozenin M C and Limoli C L 2019 Long-term neurocognitive benefits of FLASH radiotherapy driven by reduced reactive oxygen species *Proceedings of the National Academy of Sciences of the United States of America* **116** 10943-51

Montay-Gruel P, Bouchet A, Jaccard M, Patin D, Serduc R, Aim W, Petersson K, Petit B, Bailat C, Bourhis J, Bräuer-Krisch E and Vozenin M C 2018 X-rays can trigger the FLASH effect: Ultra-high dose-rate synchrotron light source prevents normal brain injury after whole brain irradiation in mice *Radiotherapy and oncology : journal of the European Society for Therapeutic Radiology and Oncology* **129** 582-8

Montay-Gruel P, Petersson K, Jaccard M, Boivin G, Germond J-F, Petit B, Doenlen R, Favaudon V, Bochud F, Bailat C, Bourhis J and Vozenin M-C 2017 Irradiation in a flash: Unique sparing of memory in mice after whole brain irradiation with dose rates above 100 Gy/s *Radiotherapy and Oncology* **124** 365-9

Paganetti H 2012 *Proton Therapy Physics* (Boca Raton: CRC Press)

Paganetti H, Blakely E, Carabe-Fernandez A, Carlson D J, Das I J, Dong L, Grosshans D, Held K D, Mohan R, Moiseenko V, Niemierko A, Stewart R D and Willers H 2019 Report of the AAPM TG-256 on the relative biological effectiveness of proton beams in radiation therapy *Medical physics* **46** e53-e78

Perl J, Shin J, Schumann J, Faddegon B and Paganetti H 2012 TOPAS: an innovative proton Monte Carlo platform for research and clinical applications *Medical physics* **39** 6818-37

Petersson K, Adrian G, Butterworth K and McMahon S J 2020 A Quantitative Analysis of the Role of Oxygen Tension in FLASH Radiation Therapy *International journal of radiation oncology, biology, physics* **107** 539-47

Smyth L M L, Donoghue J F, Ventura J A, Livingstone J, Bailey T, Day L R J, Crosbie J C and Rogers P A W 2018 Comparative toxicity of synchrotron and conventional radiation therapy based on total and partial body irradiation in a murine model *Scientific Reports* **8** 12044

Usman S and Patil A 2018 Radiation detector deadtime and pile up: A review of the status of science *Nuclear Engineering and Technology* **50** 1006-16

Vozenin M C, De Fornel P, Petersson K, Favaudon V, Jaccard M, Germond J F, Petit B, Burki M, Ferrand G, Patin D, Bouchaab H, Ozsahin M, Bochud F, Bailat C, Devauchelle P and Bourhis J 2019a The Advantage of FLASH Radiotherapy Confirmed in Mini-pig and Cat-cancer Patients *Clinical cancer research : an official journal of the American Association for Cancer Research* **25** 35-42

Vozenin M C, Hendry J H and Limoli C L 2019b Biological Benefits of Ultra-high Dose Rate FLASH Radiotherapy: Sleeping Beauty Awoken *Clin Oncol (R Coll Radiol)* **31** 407-15

Wilson J D, Hammond E M, Higgins G S and Petersson K 2019 Ultra-High Dose Rate (FLASH) Radiotherapy: Silver Bullet or Fool's Gold? *Front Oncol* **9** 1563

Yang Y, Shi C, Chen C-C, Tsai P, Kang M, Huang S, Lin C-H, Chang F-X, Chhabra A M, Choi J I, Tome W A, Simone C B and Lin H A 2D strip ionization chamber array with high spatiotemporal resolution for proton pencil beam scanning FLASH radiotherapy *Medical physics* **49**

Zhou S, Rao W, Chen Q, Tan Y, Smith W, Sun B, Zhou J, Chang C-W, Lin L, Darafsheh A, Zhao T and Zhang T 2022 A multi-layer strip ionization chamber (MLSIC) device for proton pencil beam scan quality assurance *Physics in Medicine & Biology* **67** 175006

Zou W, Diffenderfer E S, Cengel K A, Kim M M, Avery S, Konzer J, Cai Y, Boisseu P, Ota K, Yin L, Wiersma R, Carlson D J, Fan Y, Busch T M, Koumenis C, Lin A, Metz J M, Teo B K and Dong L 2021 Current delivery limitations of proton PBS for FLASH *Radiotherapy and oncology : journal of the European Society for Therapeutic Radiology and Oncology* **155** 212-8